\documentclass{SciPost}

\hypersetup{
    colorlinks,
    linkcolor={red!50!black},
    citecolor={blue!50!black},
    urlcolor={blue!80!black}
}

\usepackage[bitstream-charter]{mathdesign}
\usepackage{scrextend}
\usepackage{amsmath}
\usepackage{graphicx}
 \usepackage{booktabs}

\usepackage{txfonts}
\usepackage{enumitem}
\usepackage{subfig}
\usepackage{placeins}
\usepackage{wrapfig}
\usepackage{caption}
\DeclareSymbolFont{usualmathcal}{OMS}{cmsy}{m}{n}
\DeclareSymbolFontAlphabet{\mathcal}{usualmathcal}

\fancypagestyle{SPstyle}{
\fancyhf{}
\lhead{\colorbox{scipostdeepblue}{\bf \color{white} ~SciPost Physics Proceedings }}
\rhead{{\bf \color{scipostdeepblue} ~Submission }}

\fancyfoot[C]{\textbf{\thepage}}
}

\begin{document}

\pagestyle{SPstyle}

\begin{center}{\Large \textbf{\color{scipostdeepblue}{
\texttt{maria} goes \texttt{NIFTy}: Gaussian Process-Based Reconstruction and Denoising of Simulated (Sub-)Millimetre Single-Dish Telescope Data\\
}}}\end{center}

\begin{center}\textbf{
J. W\"urzinger\textsuperscript{1,2$\star$},
J.~van Marrewijk\textsuperscript{3},
T.~W.~Morris\textsuperscript{4,5},
R.~Fuchs\textsuperscript{1},
T.~Mroczkowski\textsuperscript{6,7} and
L.~Heinrich\textsuperscript{1,2}
}\end{center}

\begin{center}
{\bf 1} Technical University of Munich, Arcisstrasse 21, 80333 Munich, Germany
\\
{\bf 2} Excellence Cluster ORIGINS, Boltzmannstrasse 2, 85748 Garching, Germany
\\
{\bf 3} Leiden Observatory, Leiden University, P.O. Box 9513, 2300 RA Leiden, The Netherlands
\\
{\bf 4} Yale University, New Haven, CT 06511, USA
\\
{\bf 5} Brookhaven National Laboratory, Upton, NY 11973, USA
\\
{\bf 6} European Southern Observatory, Karl-Schwarzschild-Str 2, Garching 85748, Germany
\\
{\bf 7} Institute of Space Sciences (ICE, CSIC), Carrer de Can Magrans, s/n, 08193 Cerdanyola del Vallès, Barcelona, Spain
\\[\baselineskip]
$\star$ \href{mailto:jonas.wuerzinger@tum.de}{\small jonas.wuerzinger@tum.de}
\end{center}

\definecolor{palegray}{gray}{0.95}
\begin{center}
\colorbox{palegray}{
  \begin{tabular}{rr}
  \begin{minipage}{0.37\textwidth}
    \includegraphics[width=60mm]{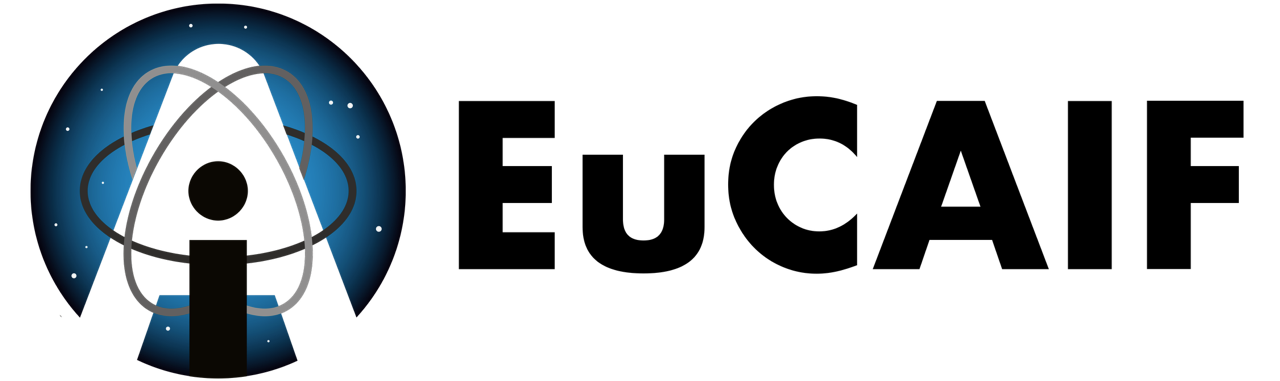}
  \end{minipage}
  &
  \begin{minipage}{0.5\textwidth}
    \vspace{5pt}
    \vspace{0.5\baselineskip} 
    \begin{center} \hspace{5pt}
    {\it The 2nd European AI for Fundamental \\Physics Conference (EuCAIFCon2025)} \\
    {\it Cagliari, Sardinia, 16-20 June 2025
    }
    \vspace{0.5\baselineskip} 
    \vspace{5pt}
    \end{center}
    
  \end{minipage}
\end{tabular}
}
\end{center}

\section*{\color{scipostdeepblue}{Abstract}}
\textbf{\boldmath{%
{(Sub-)millimetre single-dish telescopes feature faster mapping speeds and access larger spatial scales than their interferometric counterparts. However, atmospheric fluctuations tend to dominate their signals and complicate recovery of the astronomical sky. Here we develop a framework for Gaussian process-based sky reconstruction and separation of the atmospheric emission from the astronomical signal based on Numerical Information Field Theory (\texttt{NIFTy}). To validate this novel approach, we use the \texttt{maria} software to generate synthetic time-ordered observational data mimicking the MUSTANG-2 bolometric array. This approach leads to significantly improved sky reconstructions versus traditional methods.
}
}}

\vspace{\baselineskip}

\noindent\textcolor{white!90!black}{%
\fbox{\parbox{0.975\linewidth}{%
\textcolor{white!40!black}{\begin{tabular}{lr}%
  \begin{minipage}{0.6\textwidth}%
    {\small Copyright attribution to authors. \newline
    This work is a submission to SciPost Phys. Proc. \newline
    License information to appear upon publication. \newline
    Publication information to appear upon publication.}
  \end{minipage} & \begin{minipage}{0.4\textwidth}
    {\small Received Date \newline Accepted Date \newline Published Date}%
  \end{minipage}
\end{tabular}}
}}
}




\section{Introduction}
\label{sec:intro}

Reconstruction of the (sub-)millimetre astronomical sky using single-dish telescope data poses several challenges. Measured time series consist of contributions from the atmosphere, astrophysical signals, the cosmic microwave background (CMB), and several sources of (in)dependent noise. Turbulent atmospheric emission often dominates over astrophysical signals by several orders of magnitude, significantly complicating traditional mapmaking methods (see e.g. \cite{morris2022, Naess2023, Morris2025, Naess2025}). These methods typically treat the atmosphere as a stochastic noise source to be filtered; however, such filtering results in a biased transfer function, particularly at large angular scales \cite{Romero2020}. This is especially problematic for the recovery of astronomical signals such as the CMB and its secondary anisotropies, which are faint, extended, and relatively time-invariant.

In this work, we aim to tackle the above challenges to the reconstruction of the signals by relying on a new approach using Bayesian Variational Inference (VI) for (sub-)mm single-dish mapmaking. For this, we use simulated (sub-)mm single-dish telescope data produced with \texttt{maria} (see~\cite{maria1})\footnote{\url{https://thomaswmorris.com/maria/}}, a versatile, python-based general-purpose simulator for mm/submm-band telescopes that generates location-specific weather including turbulence in an evolving atmosphere. 
We then apply reconstruction and denoising techniques we developed (publicly available)\footnote{\url{https://github.com/jwuerzinger/CMB_denoising/}} based on the Bayesian Numerical Information Field Theory (\texttt{NIFTy}) framework (see \cite{niftyre})\footnote{\url{https://ift.pages.mpcdf.de/nifty/}}, which has successfully been used in reconstruction for imaging and many other fields in astrophysics (e.g.\ \cite{scheel2023, edenhofer2024, roth2024, westerkamp2024}). 

\section{Methodology}

Our methodology involves a two-tiered approach. First, we use \texttt{maria} to produce synthetic observational data tailored to mimic observations from the second-generation Multiplexed SQUID/TES Array at Ninety Gigahertz (MUSTANG-2; see \cite{Dicker2014}), a 223-element bolometer detector array on the 100-meter Green Bank Telescope in West Virginia, USA.  
Following \cite{maria1}, we perform a 30 minute simulated MUSTANG-2 observation of the thermal Sunyaev-Zeldovich effect from a galaxy cluster, a scattering of the CMB that traces the energetics of the hot thermal gas (see \cite{Mroczkowski2019, DiMascolo2025} for details). We adopt a sampling rate of $50\,\mathrm{Hz}$.

The second tier of our approach is to apply the \texttt{NIFTy} framework to denoise the time-ordered data (TODs) and jointly reconstruct the astronomical signal and the time series contributions from the atmosphere. In this framework, signals are modelled by a Gaussian process (GP) based ``Correlated Field Model'' (CFM), which describes correlations between nearby pixels using a learnable power spectrum in Fourier space. For more details on the CFM, refer to Ref.\ \cite{arras2022}.

We employ the CFM to describe different contributions in the measurement equation,
\begin{equation}
    d(t)=R(s,t)+a(t)+n,
\end{equation}
where the time series data $d(t)$ are given by a non-linear time-dependent signal response function $R(s,t)$, an additional atmospheric component $a(t)$, and noise $n$. More specifically, $R$ implements a scanning trajectory over the signal map $s$ after beam convolution. The scan strategy adapted is the Lissajous daisy pattern found to be optimal for MUSTANG-2 observations (see \cite{Romero2020,maria1}). 
The time-independent signal map $s$ and the time-dependent atmosphere $a(t)$ are modelled with a two/one-dimensional CFM, respectively. We treat the map $s$ as constant in time and hence intrinsically two-dimensional. As shown in Refs.\ \cite{morris2022, maria1}, the atmosphere $a(t)$ is largely correlated across the detectors and can hence be modelled as one-dimensional, assuming the same response for all detectors.

The noise $n$ is modelled as an uncorrelated Gaussian component over which we marginalise.
As the frequency of atmospheric fluctuations is much lower than the simulated observation frequency of $50\,\mathrm{Hz}$, the atmosphere response is downsampled by a factor of three, meaning that one time-step in the atmosphere GP models three simulated atmosphere time-steps.

Together, the signal map and atmosphere CFMs comprise our forward model,
\begin{equation}
    f(\xi) = R(s(\xi_{s})) + a(\xi_{a}),
\end{equation}
with model parameters $\xi = (\xi_{s}, \xi_{a})$, 
where $\xi_{s}$ and $\xi_a$ are parameters modelling the power-spectrum and the pixel-by-pixel excitations in the spatial/time directions for the respective signals.
Assuming independent Gaussian noise with diagonal covariance $N$, the likelihood is given by
\begin{equation}
    p(d\,|\,\xi) = \mathcal{G}(d - f(\xi)\,|\,N).
\end{equation}

In order to approximate the posterior distribution $p(\xi\,|\,d)$, \texttt{NIFTy} relies on geometric Variational Inference (geoVI) \cite{frank2021}. By exploiting the geometric properties of the posterior, geoVI allows us to iteratively optimise a set of posterior samples for complex Bayesian models with millions of parameters. From these samples, we can derive posterior estimates and uncertainties for quantities of interest like the signal map and the atmospheric contributions.

In total, the CFM used in this work constitutes about $19$ million data points, which are used to model about $1.1$ million degrees of freedom, resulting in a relatively expensive reconstruction process that takes around $2.5$ hours on one core of an Intel\textcopyright\  Xeon\textcopyright\  Platinum 8468V processor and one NVIDIA\textcopyright\  H100 GPU.

\begin{figure}[h!]
    \centering
    \includegraphics[clip,trim=5mm 4mm 0mm 10mm,width=0.95\linewidth]{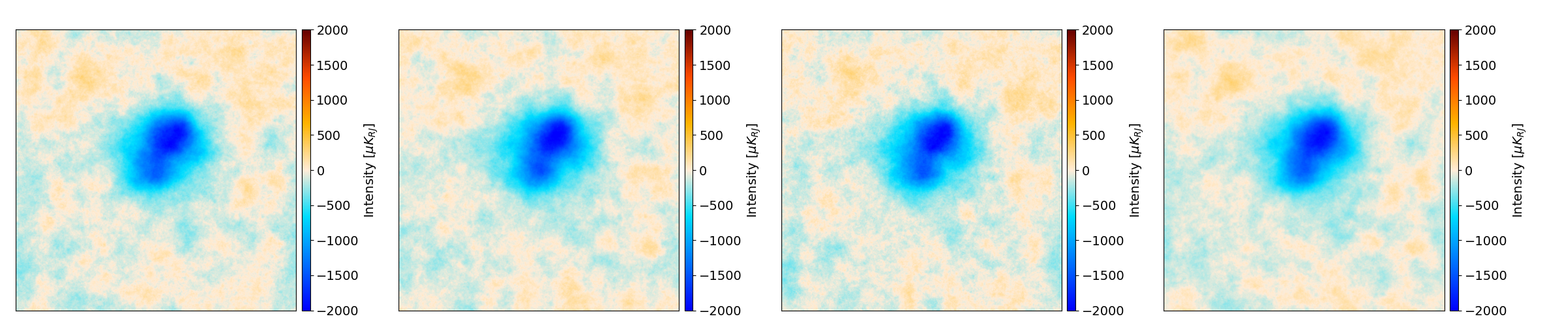}
    \caption{Output samples from our Bayesian reconstruction method, showing the variation in the \texttt{NIFTy} sky reconstructions through VI. Each panel is 5$^\prime$ across.
    For comparison, the ground truth map can be seen in the upper right panel of Fig.\ \ref{fig:final_map_comp}, while the mean \texttt{NIFTy} reconstruction and residuals are in the lower left and central panels of the figure.}
    \label{fig:final_map_samples}
\end{figure}

\begin{figure}[h!]
    \centering
    \includegraphics[clip,trim=3mm 4mm 1mm 0mm, width=0.95\textwidth]{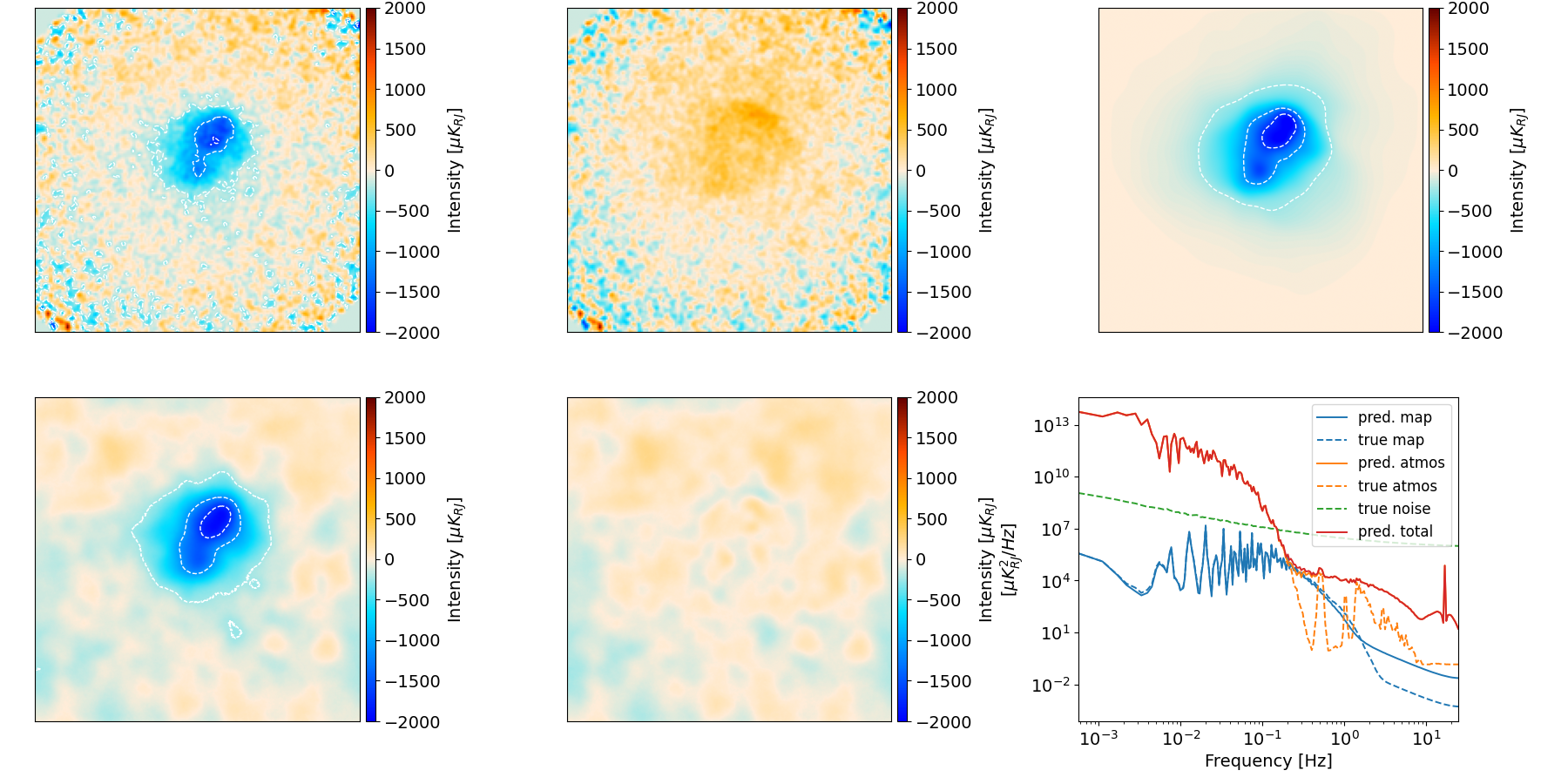}
    \caption{Compilation of the final maps and power spectral densities. 
    The upper left shows the maximum likelihood map produced by mapping synthetic data from \texttt{maria} using \texttt{minkasi} \cite{minkasi}. The upper middle shows the residuals on this reconstruction after subtracting the beam-smoothed input map (upper right panel). Each map panel is 5$^\prime$ across.
    The lower left and lower central panels show, respectively, the \texttt{NIFTy} reconstruction and residuals.
    The final, lower right panel shows the power spectral density of the data and fit components. 
    In particular, the true and \texttt{NIFTy}-predicted map TODs closely follow each other from $\approx$2~Hz, where the array noise dominates, down to $\lesssim 1$~mHz, which for a typical scan speed of $50''$~s$^{-1}$ corresponds to scales $\gtrsim 13.9^{\circ}$. 
    }
    \label{fig:final_map_comp}
\end{figure}


\section{Results}

After optimisation, we obtain four posterior samples of possible map realisations, which are shown in Fig.~\ref{fig:final_map_samples}. From this ensemble, we compute both the standard deviation, which provides an estimate of the reconstruction uncertainty on our astronomical signal, and the mean across the reconstructions. The reconstructed maps have a higher on-source time in the centre due to the adopted scanning strategy, resulting in a better reconstruction with lower variance in the centre and higher sample variance/uncertainty in the outskirts.

To validate our approach, we compare the results with those produced using the maximum likelihood (ML) mapmaking method \texttt{minkasi} \cite{Romero2020, minkasi} (see Fig.~\ref{fig:final_map_comp}).
In comparison to the ML map, we find that the \texttt{NIFTy} reconstruction is quantitatively closer to the ground truth map and exhibits lower noise levels, particularly on small scales. The maximum residual for the \texttt{NIFTy} reconstruction is reduced from about \(1920\,\mu\mathrm{K}_\mathrm{RJ}\) (\(\sim96\%\)) to about \(250\,\mu\mathrm{K}_\mathrm{RJ}\) (\(\sim13\%\)) and appears more uniform on a whole and not strongly correlated with the astronomical signal of interest.
The residual of the respective reconstructions is shown in the middle column of Fig.~\ref{fig:final_map_comp}. Here, the ML reconstruction shows an average residual offset of \(\sim 70\,\mu\mathrm{K}_{\mathrm{RJ}}\), likely due to filtering, while the \texttt{NIFTy} reconstruction remains fiducial to within \(1.5\,\mu\mathrm{K}_{\mathrm{RJ}}\) of the ground truth on average.

The power spectra of all reconstructed signals and their true (simulated) counterparts are shown in the lower right panel of Figure \ref{fig:final_map_comp}. Both the atmosphere and the signal map are accurately reconstructed over many orders of magnitude in both power and frequency, despite a noise contribution which is roughly three orders of magnitude larger than the signal from the map. The map reconstruction breaks down at a frequency of about $1\,\mathrm{Hz}$. In this regime, the frequency-dependent $1/f$ noise component (see \cite{maria1} for more detailed discussion) becomes significant and, since it is currently not included in the \texttt{NIFTy} model, cannot be subtracted and thus enters the map. This effect can be seen in the map shown in Figure \ref{fig:final_map_comp} as well.
We note that modelling this $1/f$ noise component is likely feasible as we are currently showing in internal tests on real data, finding that multiple noise models are required as the behaviour is not perfectly correlated across the detectors. The results will be published in a future work (Fuchs et al.\ in prep.).

\section{Conclusion \& Outlook}

We introduced a Bayesian VI framework--specifically, the Numerical Information Field Theory (\texttt{NIFTy})--for Gaussian process (GP)-based atmospheric emission removal and reconstruction of astronomical signals from single-dish (sub-)mm data. 
We find that the GP-based reconstruction can accurately separate the atmospheric from the astronomical components down to levels comparable to the detector array--averaged noise, while also providing posterior reconstruction samples that can be used to define both a reconstruction mean and its associated uncertainty.

The next area of development will be to apply these methods to observations from MUSTANG-2 and other real instruments, using their data to improve modelling of the instrumental pink-noise ($1/f$) component. Beyond this, we are working toward improving scalability for mapping larger datasets, including longer observations, larger detector arrays, and multi-banded or multi-frequency observations, in order to test scale recovery for faint, extended signals such as the CMB. This will be particularly relevant for mapping data from, e.g., the Simons Observatory (SO; \cite{solatscience2025}) as well as the Atacama Large Aperture Submillimeter Telescope (AtLAST), a concept for a next-generation 50-meter telescope covering 30--950~GHz \cite{Mroczkowski2025}. This effort also includes parallelisation to utilise multiple GPU cards, both for speed improvements and to provide sufficient memory for larger datasets.

\section*{Acknowledgements}
We thank Luca Di Mascolo, Mats Kirkaune, Adrian Duivenvoorden and Jakob Knollm\"uller for valuable input and feedback throughout this work.
We are grateful for the support of the Origins Data Science Lab (ODSL) team, and to Nicole Hartman in particular.  ODSL is supported by the Excellence Cluster ORIGINS and the Deutsche Forschungsgemeinschaft as noted below.

\paragraph{Funding information}
This project is funded by the Deutsche Forschungsgemeinschaft (DFG, the German Research Foundation) under Germany's Excellence Strategy – EXC 2094 – 390783311.
TM acknowledges support from the Agencia Estatal de Investigaci\'on (AEI) and the Ministerio de Ciencia, Innovaci\'on y Universidades (MICIU) Grant ATRAE2024-154740 funded by MICIU/AEI//10.13039/501100011033.
TM is  partly supported by the program Unidad de Excelencia María de Maeztu CEX2020-001058-M, financed by MCIN/AEI//10.13039/501100011033, and by the MaX-CSIC Excellence Award MaX4-SOMMA-ICE.







\bibliography{maria, nifty}


\end{document}